\documentclass[prd,twocolumn,amsmath,amssymb,floatfix,superscriptaddress]{revtex4-1}
\usepackage{bm}
\usepackage{amsmath}
\usepackage{epsfig}
\usepackage{color}
\usepackage{natbib}
\usepackage{graphicx}
\usepackage{hyperref}
\usepackage{ifthen}
\usepackage{xstring}

\newcommand{\Mpch}{h^{-1} \mbox{Mpc}}

\newcommand{\Msunh}{h^{-1} M_\odot}
\newcommand{\Mv}{M_{\rm v}}
\newcommand{\sigmafr}{\sigma_{f_R}}
\newcommand{\sigmagr}{\sigma_{\Lambda \rm CDM}}

\def\mnras{Mon.\ Not.\ R.\ Astron.\ Soc.}
\definecolor{darkgreen}{cmyk}{0.85,0.2,1.00,0.2} 
\definecolor{purple}{cmyk}{0.5,1.0,0,0}

\begin{document}
\title{Chameleon Halo Modeling in \boldmath$f(R)$ Gravity}

\author{Yin Li}

\affiliation{Kavli Institute for Cosmological Physics, University of Chicago, Chicago Illinois 60637}
\affiliation{Department of Physics, University of Chicago, Chicago IL 60637}

\author{Wayne Hu}
\affiliation{Kavli Institute for Cosmological Physics, University of Chicago, Chicago Illinois 60637}
\affiliation{Department of Astronomy and Astrophysics, University of Chicago, Chicago Illinois 60637}

\date{\today}

\begin{abstract}
\baselineskip 11pt
We model the chameleon effect on cosmological statistics for the modified
gravity $f(R)$ model of cosmic acceleration.   The chameleon effect, required to make the model 
compatible with local tests of gravity, reduces force
enhancement as a function of the depth of the gravitational potential wells of collapsed
structure and so is readily incorporated into a halo model by including parameters for the chameleon mass threshold and rapidity of transition.   We show that the abundance
of halos around the chameleon mass threshold is enhanced by both the merging from below
and the lack of merging to larger masses.   
This property also controls the power spectrum in the nonlinear regime and we provide a description of the transition to the linear regime
that is valid for a wide range of $f(R)$ models.
 \end{abstract}

\maketitle

%%%%%%%%%%%%%%%%%%%%%%%%%%%%%%%%%%%%%%%%%
\section{Introduction} \label{sec:intro}
%%%%%%%%%%%%%%%%%%%%%%%%%%%%%%%%%%%%%%%%%

The modified action $f(R)$ model for cosmic acceleration provides a concrete framework
under which to relate local and cosmological tests of gravity.  
Here the Einstein-Hilbert action is augmented with a general function of the Ricci scalar
curvature
in such a way as to mimic the cosmological constant at cosmologically
low curvature \cite{Capozziello:2002rd,NojOdi03,Caretal03}.

The extra scalar degree of freedom $d f/dR$ mediates an enhanced gravitational 
force on scales smaller than its Compton wavelength.
In order to hide this enhancement from local tests of gravity, viable models
employ the chameleon mechanism \cite{Mota:2003tc,khoury04a} where the 
Compton wavelength can shrink in regions of deep gravitational potential
wells.   If one assumes that the $f(R)$ model is a complete description of
gravity from solar system scales to cosmology, local tests in conjunction with 
minimal assumptions about the galactic potential 
 place a stringent
bound on $f(R)$ models.   Namely the cosmological amplitude of the field $|f_{R0}|$ must
be less than a few times $10^{-6}$ \cite{HuSaw07a}.

Currently the tightest cosmological bounds come from the lack of an excess in
the abundance of the most massive dark matter halos or galaxy clusters
\cite{Schmidt:2009am,Lombriser:2010mp}.  The bound on the field amplitude
depends on the model but generally lies above the $10^{-5}$ level for models that seek
to diminish the modification at high redshift \cite{Ferraro:2010gh}.
Thus cosmological constraints currently are at best approaching the level of
solar system tests.  On the other hand they test these modifications on a vastly different
scale and it  is possible that $f(R)$ is just an effective theory only strictly valid on cosmological scales.

As cosmological data continue to improve, they will soon begin to probe
a regime where cosmology is directly competitive with solar system tests in the full
$f(R)$ context. 
Here the chameleon mechanism becomes important for all halos of mass comparable
to or greater than the Galaxy.    This regime is substantially more difficult to characterize
 due to the non-linearity in the field equation for the enhanced force.  

Under the halo model approach, the first step in understanding cosmological
statistics associated with the observable properties of dark matter halos is to
characterize the mass function.
Previous attempts to model the mass function of simulations have not been able to
capture the abundance of intermediate mass halos once the chameleon mechanism becomes
active \cite{Halopaper}.    This abundance is doubly enhanced since 
 the extra force augments merging of  low mass halos into intermediate mass halos
 whereas the chameleon effect shuts down merging of intermediate mass halos to high mass halos.

Based on a mass function model, we can build a  
parameterized post-Friedmann (PPF) description of the chameleon effect in cosmological 
statistics such as the dark matter power spectrum.  
Whereas previous PPF approaches have been based on 
density thresholds  \cite{HuSaw07b} 
and are appropriate for models that utilize the Vainshtein
mechanism, they require extensive ad hoc modifications for the chameleon mechanism \cite{Koyama:2009me}.
The problem is that the chameleon mechanism should be 
parameterized in terms of a proxy for the depth of  gravitational potential wells,
which in the halo model can be readily associated with the mass of dark matter halos.

The outline of the paper is as follows.  In \S \ref{sec:frsims}, we briefly review the $f(R)$ model
and simulations.  We parameterize the chameleon effect in the mass function in 
\S \ref{sec:massfunction}, which we use in \S \ref{sec:powerspectrum} to characterize the
nonlinear dark matter power spectrum.   We discuss these results in \S \ref{sec:discussion}.

%%%%%%%%%%%%%%%%%%%%%%%%%%%%%%%%%%%%%%%%%
\section{$f(R)$ Simulations}
\label{sec:frsims}

In  $f(R)$ models, the Einstein-Hilbert action is augmented with a general function of the scalar curvature $R$
 \begin{eqnarray}
S_{G}  =  \int{d^4 x \sqrt{-g} \left[ \frac{R+f(R)}{16\pi G}\right]}\,. 
\label{eqn:action}
\end{eqnarray}
Here and throughout $c=\hbar=1$.    
For definiteness we take the high curvature limit of the models in \cite{HuSaw07a}
\begin{eqnarray}
%f(R) = -16 \pi G \rho_\Lambda - f_{R0} \frac{\bar R_0^2}{ R} \,,
f(R) \approx -2 \Lambda -\frac{ f_{R0}}{n} \frac{\bar R_0^{n+1}}{ R^n} \,,
\label{eqn:fRapprox}
\end{eqnarray}
where the constant $\bar R_0$ is the background scalar curvature $R$ today in a $\Lambda$CDM cosmology
with a cosmological constant $\Lambda$.  Here $f_{R0}$ is a parameter that controls
the amplitude of the field $f_{R}\equiv df/dR$ at the background curvature today.  Note that viable models have
$|f_{R0}| \ll1$ and expansion histories that are observationally indistinguishable from
$\Lambda$CDM.

Gravitational force enhancements are associated with the field $f_R$
whose fluctuations from the background $\delta f_R = f_R(R) - f_R(\bar R)$ obey a non-linear Poisson equation in comoving coordinates
\begin{eqnarray}
\nabla^2 \delta  f_{R} = \frac{a^{2}}{3}\left[\delta R(f_{R}) - 8 \pi G \delta \rho_{\rm m}\right] \,.\label{eqn:frorig}
\end{eqnarray}
These field fluctuations act an additional source to the gravitational potential
\begin{eqnarray}
\nabla^2 \Psi = {4\pi G}a^{2} \delta \rho_{\rm m} - \frac{1}{2} \nabla^2 \delta  f_{R} \,,\label{eqn:potorig}
\end{eqnarray}
whose gradients accelerate non-relativisitic particles as usual.
These equations have been solved with N-body techniques for $n=1$, $n=2$
and a variety of amplitudes $f_{R0}$ 
with cosmological parameters
$\Omega_{\rm m}=1-\Omega_\Lambda=0.24$,
 $h=0.73$ and an initial power 
spectrum with $A_s = (4.89\times 10^{-5})^2$ at $k=0.05$Mpc$^{-1}$ and
$n_s=0.958$
\cite{Pkpaper,Halopaper,Ferraro:2010gh} \footnote{This corrects a 3\% absolute
error in the initial amplitude reported in \cite{Pkpaper} but does not change results
relative to $\Lambda$CDM significantly.}.

The nonlinearity of Eq.~(\ref{eqn:frorig}) is the key to understanding when the chameleon
mechanism does and does not operate.
If the field fluctuations are small, they can be linearized as 
\begin{equation}
\delta R \approx { {d R \over d f_{R}} \bigg|_{\bar R} \delta f_R } = 3 \lambda_C^{-2} a^{-2} \delta f_R \,,
\end{equation}
where $\lambda_C$ is the comoving Compton wavelength of the $f_R$ field
in the background.   In this case the Poisson equation~(\ref{eqn:potorig}) has the Fourier 
solution
\begin{equation}
k^2 \Psi = -4 \pi G \left( {4 \over 3} - {1 \over 3} {1 \over k^2\lambda_C^2 + 1} \right) a^2  \delta \rho_{\rm m}\,,
\label{eqn:nochameleon}
\end{equation}
and gravitational forces are enhanced by $1/3$ on scales that are below $\lambda_C$. 
We call this the ``no-chameleon" limit and for comparison to the chameleon simulations, 
runs with the same initial conditions using Eq.~(\ref{eqn:nochameleon}) were conducted.
In this limit field fluctuations follow the local gravitational potential on small scales.

As  local gravitational potentials of dark matter halos  deepen, the field fluctuation can become comparable
to the background value $f_{R0}$. The Compton 
wavelength can then change significantly from its background value.   For the $f(R)$ models
described by Eq.~(\ref{eqn:fRapprox}) the Compton wavelength shrinks so that the
force enhancement disappears in deep gravitational potential wells.   This ability  is
called the chameleon mechanism in the literature \cite{Mota:2003tc,khoury04a}.  

Finally,  $\Lambda$CDM models with the same initial conditions were simulated for comparison.   Thus for each simulation box size  ($64, 128, 256, 400$ $\Mpch$) and initial conditions realization (up to 6 each), there are three simulation types: the full $f(R)$ or chameleon, the no-chameleon, and the $\Lambda$CDM runs.

To identify dark matter halos in the simulations, we use a spherical overdensity algorithm  centered around local density peaks  similar to \cite{Tin08}.  This method differs slightly from the center of mass of the whole halo approach  of \cite{LacCol94} used
in \cite{Halopaper} and is thought to be more directly related to halo observables.

Briefly, to make a crude sorting of density peaks we first assign particles to the grid
using the cloud-in-cell (CIC) scheme.
 Starting at the highest density grid point, we grow a halo until the enclosed spherical overdensity reaches 
$\Delta = \rho_m/\bar\rho_m = 200$ defining a radius $r_{200}$.   
We then refine the center of the halo by solving for the center of mass iteratively in shrinking radii from $r_{200}/3$ to
$r_{200}/15$ or until only 20 particles are found within the smoothing radius.
We then regrow the halo around this center until the spherical overdensity criterion is met.
The halo mass $M_{200}$ is the sum of mass of all particles 
enclosed within this halo radius $r_{200}$.  Due to lack of spatial resolution 
 we only count halos with more than $N_{\rm min} = 800$ particles.
 
 Compared with the halo center of mass approach, this algorithm tends to break up
 regions with large sub-clumps into separate halos reducing the number of
 high mass halos  by up to 10\%.   Its effect on the relative abundance between the $f(R)$ simulations and
 the $\Lambda$CDM simulations is much smaller since it affects the two  in the same way.
 
 We construct the mass function of dark matter halos for
the various simulations by adding up the number of halos within a certain mass bin from different boxes directly, which implicitly weights the simulations by volume.  To estimate the
errors on the mass function enhancements over $\Lambda$CDM, we bootstrap resample the differences, 
with replacement within
each of the different box sizes.

\begin{figure*}[tbph]
\centering
\includegraphics[width=0.95\textwidth]{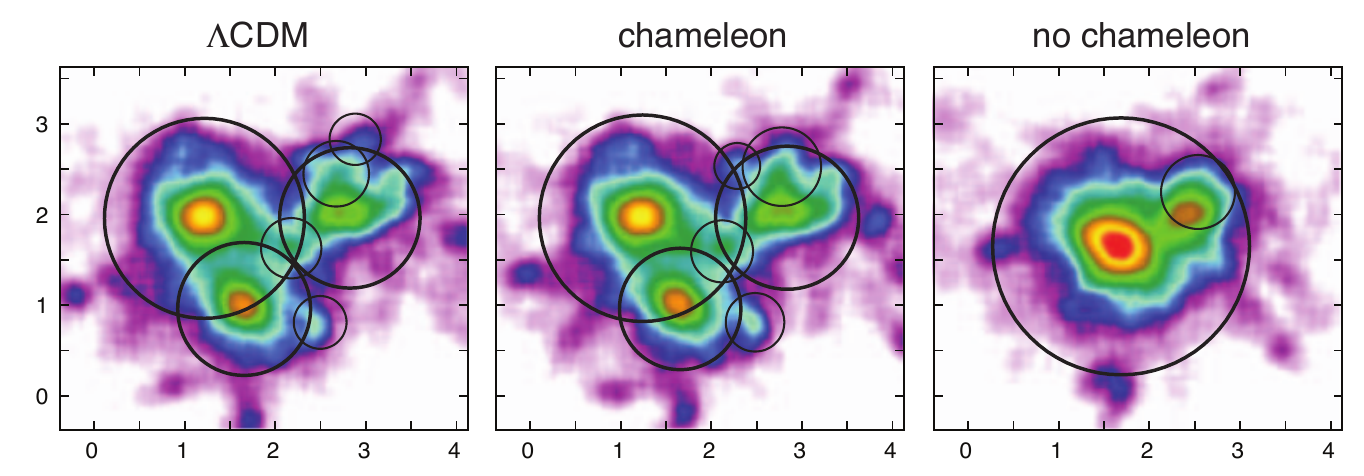}
\caption{Projected 2D density maps and identified halos for the 3 types of simulations: $\Lambda$CDM, $f(R)$ chameleon,
and no-chameleon models $(|f_{R0}|=10^{-6}, n=1)$.  Axes are in $\Mpch$ and the color map represents the logarithmic  density on a
$0.25 \Mpch$ grid projected across a $4.25 \Mpch$ depth. 
Halos are denoted with circles (thick lines: $>800$ particles; thin lines $100-800$ particles
plotted for reference).
The largest halos in each case from left to right are $7.5\times10^{13}$,
$8.2\times10^{13}$ and $1.6\times10^{14}\Msunh$ respectively showing that between $\Lambda$CDM and the
chameleon case halos mainly grow in mass whereas between the chameleon and no-chameleon case a major merger has
occurred.}
\label{fig:merge}
\end{figure*}

\section{Chameleon Mass Function}
\label{sec:massfunction}

The mass function or differential abundance of dark matter halos for the
various $f(R)$ models and simulations has previously been studied in \cite{Halopaper,Ferraro:2010gh}.
In the large field regime of $|f_{R0}| > 10^{-5}$, the excess abundance appears
mainly in the rarest halos and these results are well-modeled
by simple modifications to spherical collapse predictions. 
In the small field regime $|f_{R0}|< 10^{-5}$ which is comparable to or
smaller than the depth of the gravitational potential wells of cluster mass sized
halos, the chameleon mechanism shuts off the excess in the abundance
of these halos.

Instead, in this regime, the chameleon simulations produce an abundance
of intermediate sized halos that is in excess of both the
no-chameleon simulations and the predictions with the full 
$1/3$ enhancement of forces everywhere \cite{Halopaper}.

To understand this result, we can compare the full and no-chameleon simulations
in the $n=1$, $|f_{R0}|=10^{-6}$ model where the chameleon effect is the
strongest.  
Since the $\Lambda$CDM, chameleon and no-chameleon models are simulated with the
same initial condition realization, we can examine regions in the simulations associated
with intermediate mass halos.  In Fig.~\ref{fig:merge} we show an example.  The halos
in the chameleon run are very similar to those in the $\Lambda$CDM run only slightly more
massive.  On the other hand, in the no-chameleon run several of the intermediate sized halos have
merged into one high mass halo.

  \begin{figure}[tbph]
\centering
\includegraphics[width=0.45\textwidth]{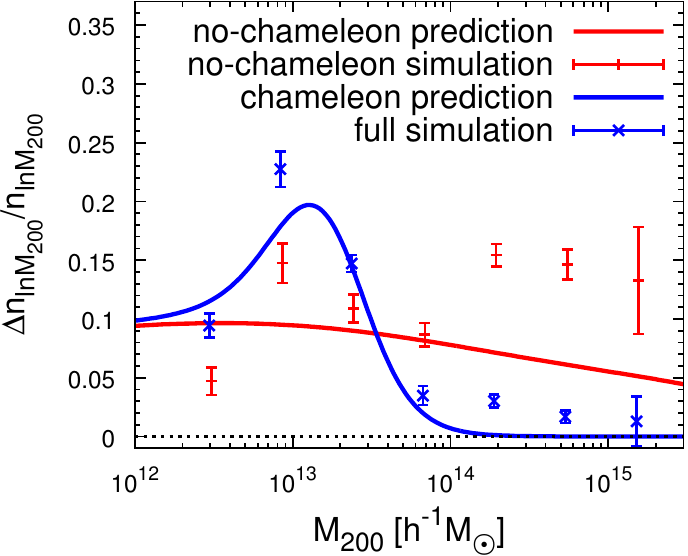}
\caption{Simulation results and parametrized post-Friedmann (PPF) fit for the mass function excess of $f(R)$ 
with $n=1$, $|f_{R0}|=10^{-6}$ over $\Lambda$CDM. Chameleon simulations show not
only a suppression of the high mass enhancement compared with no-chameleon runs
 but also a larger excess at intermediate masses, here $\sim 10^{13} h^{-1} M_\odot.$  These
 features can be fit by an interpolation between the limiting $\sigma(M)$ cases
 across a threshold mass $M_{\rm th} = 1.345\times 10^{13} h^{-1} M_\odot$ with
 $\alpha = 2.448$ for the rapidity of the transition (see Eq.~\ref{eqn:sigmamform}).
}
\label{fig:fit_n1e6}
\end{figure}

Correspondingly, 
the no-chameleon simulations show an excess in the
abundance of high mass halos which compensates the excess of intermediate
mass halos in the full runs (see Fig.~\ref{fig:fit_n1e6}).  
Those halos that under the no-chameleon assumption would have merged
to form high mass halos no longer do in the chameleon simulations causing a pileup
effect at intermediate masses.

The results are consistent with
mass conservation in the intermediate to high mass halo regime.   
To characterize this merging effect, we model the mass function based
on the Press-Schechter ansatz that all of the mass in the
universe is in halos of some mass.
We automatically conserve mass in halos
if we only vary the ingredients of the models.

The standard prescription based on the linear power spectrum requires modification however.  
In this prescription 
large masses are related to large scales through 
\begin{equation}
\sigma^2(M(R)) = \int {d^3 k \over (2\pi)^3 } | W(kR)|^2 P_{L}(k) \,,
\label{eqn:sigmamlinear}
\end{equation}
the variance of the linear
density field convolved with a tophat of radius $R$, with Fourier transform $W(kR)$,
 that encloses the mass $M$ at the background
density.   On the other hand, the 
 chameleon
mechanism operates on large masses and {\it  small} scales. 
Our approach is to retain a mass function construction based on a $\sigma(M)$ but 
generalize its relationship to the linear power spectrum in Eq.~(\ref{eqn:sigmamlinear})
 such that it no longer represents 
the rms of the linear density field of the $f(R)$ model.  We follow the parameterized post-Friedmann (PPF)  approach of taking this generalization to be an interpolation between modified and unmodified gravity
\cite{HuSaw07b}.

We take the mass function to be universal in the
virial mass $\Mv$
\begin{equation}
n_{\ln \Mv} \equiv {d n \over d\ln \Mv} = {\bar\rho_m \over  \Mv} {d \ln \nu \over d\ln \Mv} \nu f(\nu)\,,
\end{equation}
where $\nu = \delta_c/\sigma(\Mv)$ and
$\int f(\nu) d\nu =1$.
For the Sheth-Tormen mass function \cite{SheTor99}
\begin{equation}
\nu f(\nu) = A \sqrt{ {2 \over \pi} a \nu^2} [1+(a\nu^2)^{-p}] \exp(-a\nu^2/2)\,,
\end{equation}
with $a=0.75$, $p=0.3$ and $A$ given by $\int d\nu f(\nu)=1$ as $A=0.3222$.
We adopt $\Delta_{\rm v}=390$ and $\delta_c=1.673$ which are the values that match
the $\Lambda$CDM predictions for $\Omega_m = 0.24$.  
Previous attempts to model the simulation results were based on adjusting
$\Delta_{\rm v}$ and $\delta_c$ in a spherical collapse motivated range 
using $\sigma(M)$ from the linear power spectrum of
the $f(R)$ model.   That technique captures the high mass end
$M > 10^{14} h^{-1} M_\odot$ for both large and small fields
but failed in the $M < 10^{14} h^{-1} M_\odot$ 
 in the small field regime.

\begin{figure}[htbp]
\centering
\includegraphics[width=0.45\textwidth]{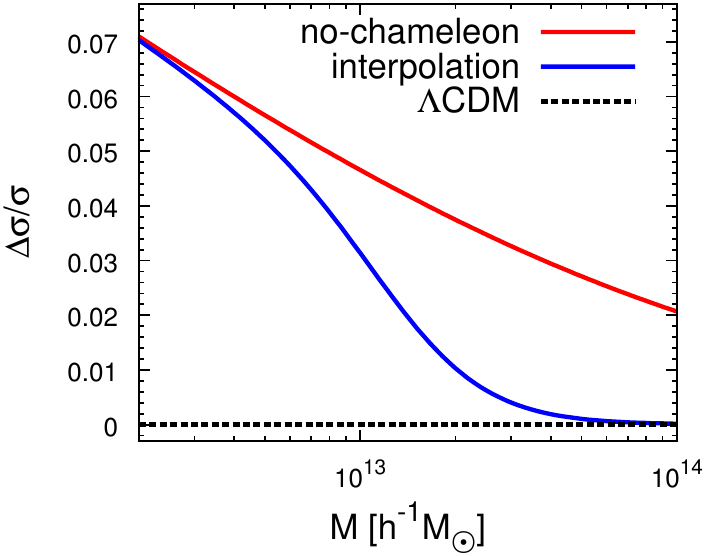}
\caption{Fractional change in $\sigma(M)$ between the linear $f(R)$ prediction (red, upper curve, $|f_{R0}|=10^{-6}$, $n=1$)
and the linear $\Lambda$CDM prediction (black, dashed curve).  
The PPF prescription interpolates between these two limits (blue, middle curve) 
with  the
transition parameters  $M_{\rm th}=1.345\times10^{13} h^{-1} M_\odot$ and $\alpha=2.448$.}
\label{fig:sigma}
\end{figure}

Here we instead leave $\Delta_{\rm v}$ and $\delta_c$ fixed but interpolate 
between  limiting behaviors of  $\sigma(M)$.    For high masses $\sigma(M)$ should approach
the $\Lambda$CDM result $\sigmagr(M)$ due to the chameleon mechanism.
For small masses, it should approach the prediction of $f(R)$ linear theory 
with enhanced forces $\sigmafr(M)$.   We thus take a chameleon PPF
  transition between
these fixed limits
\begin{equation}
\sigma(M) = { \sigmafr(M)  + (M/M_{\rm th})^\alpha \sigmagr(M) 
\over 1+  (M/M_{\rm th})^\alpha }\,.
\label{eqn:sigmamform}
\end{equation}
In Fig.~\ref{fig:sigma}, we show an example for the $|f_{R0}|=10^{-6}$, $n=1$ model.
By definition, the fraction of the universe tied up in halos above an $M \ll M_{\rm th}$ is
conserved independently of the transition
\begin{eqnarray}
F(>M) &=& \int_{M}^{\infty} {d n \over d\ln {\Mv}}  {{\Mv} \over \bar\rho_m} d\ln {\Mv} 
 \nonumber\\&=& 
\int_{{\delta_c / \sigma(M)}}^\infty d \nu f(\nu) 
\nonumber\\&\approx &
\int_{{\delta_c / \sigmafr(M)}}^\infty d \nu f(\nu), \quad M \ll M_{\rm th}\,.
\end{eqnarray}

Given this  ($M_{\rm th},\alpha$) parameterization, we fit the simulation results of
the $|f_{R0}|=10^{-6}$, $n=1$ model.  
We first convert the parameterized mass function defined at the
virial overdensity to $M_{200}$ assuming a Navarro-Frenk-White (NFW) profile
\cite{HuKravtsov}.   We then integrate the resulting $n_{\ln M_{200}}$ over a tophat in $\ln M_{200}$ of the bin size which effectively smooths
the predictions to our simulation binning.   Finally, using the bootstrap errors we minimize
the $\chi^2$ between the model and the simulation data and determine that
$M_{\rm th}= 1.345 \times 10^{13} h^{-1} M_\odot$ and $\alpha = 2.448$ yield the best fit
(see Fig.~\ref{fig:fit_n1e6}).  Given the approximate nature of the bootstrap errors, these parameters provide a reasonable fit to the simulation data.   

\begin{figure}[htbp]
\centering
\includegraphics[width=0.45\textwidth]{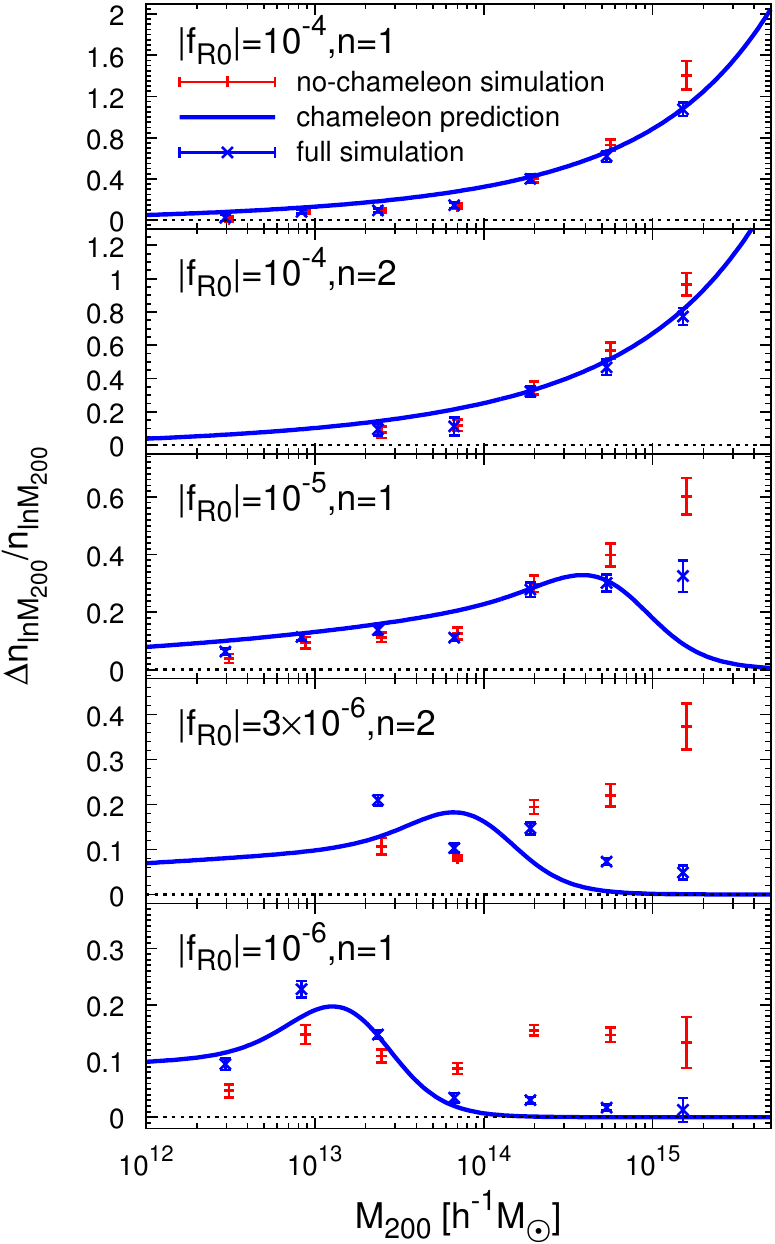}
\caption{Chameleon mass function excess for different $f(R)$ models.  A universal fit to
$\sigma(M)$  with $\alpha = 2.448$ and $M_{\rm th} = 1.345\times 10^{13} ( |f_{R0}| / 10^{-6} )^{3/2}\Msunh$ fits the range of ($|f_{R0}|$, $n$) simulations comparably well.}
\label{fig:multi}
\end{figure}

This single choice of parameters ($M_{\rm th},\alpha$) can be scaled to fit all of the simulations without introducing any additional degrees of freedom.
Given that the critical mass for the chameleon scales with the background field value 
as $|f_{R0}|^{3/2}$ \cite{Ferraro:2010gh}, we take
\begin{equation}
M_{\rm th} = 1.345\times10^{13}   \left( { | f_{R0} | \over 10^{-6}} \right)^{3/2}h^{-1}  M_\odot \,.
\end{equation}
In Fig.~\ref{fig:multi} we compare the results of the various ($|f_{R0}|$, $n$)  simulations to
this universal scaling.

\section{Chameleon Power Spectrum}
\label{sec:powerspectrum}

The chameleon mass function is the starting point for halo modeling of cosmological
observables.   For example, under the halo model, the dark matter power spectrum
is described by density correlations within halos and between halos integrated over the
mass function.   For the same wave number $k$, the chameleon mechanism
affects density correlations associated with some but not all dark matter halos.

Under the halo model, the  power spectrum in the deeply nonlinear regime is determined
by  density correlations
within single halos.   
The power spectrum in this regime can be modeled with the one halo term
\begin{equation}
P_H(k) = \int d\ln M_{\rm v} n_{\ln M_{\rm v}}\left( {M_{\rm v} \over \rho_m} \right)^2 | y(k,M_{\rm v}) |^2 \,,
\end{equation} 
where $y(k,M)$ is the Fourier transform of the density profile truncated at $r_{\rm v}$.   
For both $\Lambda$CDM and $f(R)$ the halo profiles are well characterized by the
NFW form \cite{Halopaper}.  We take a concentration given by \cite{Buletal01}
\begin{equation}
c = 9 (M_{\rm v}/M_*)^{-0.13}\,,
\end{equation}
where $M_*$ is defined via $\sigma(M_*)=\delta_c$.  Thus the main difference in this regime for
the power spectra of the models should come from the difference between the mass functions.

\begin{figure}[tbph]
\centering
\includegraphics[width=0.5\textwidth]{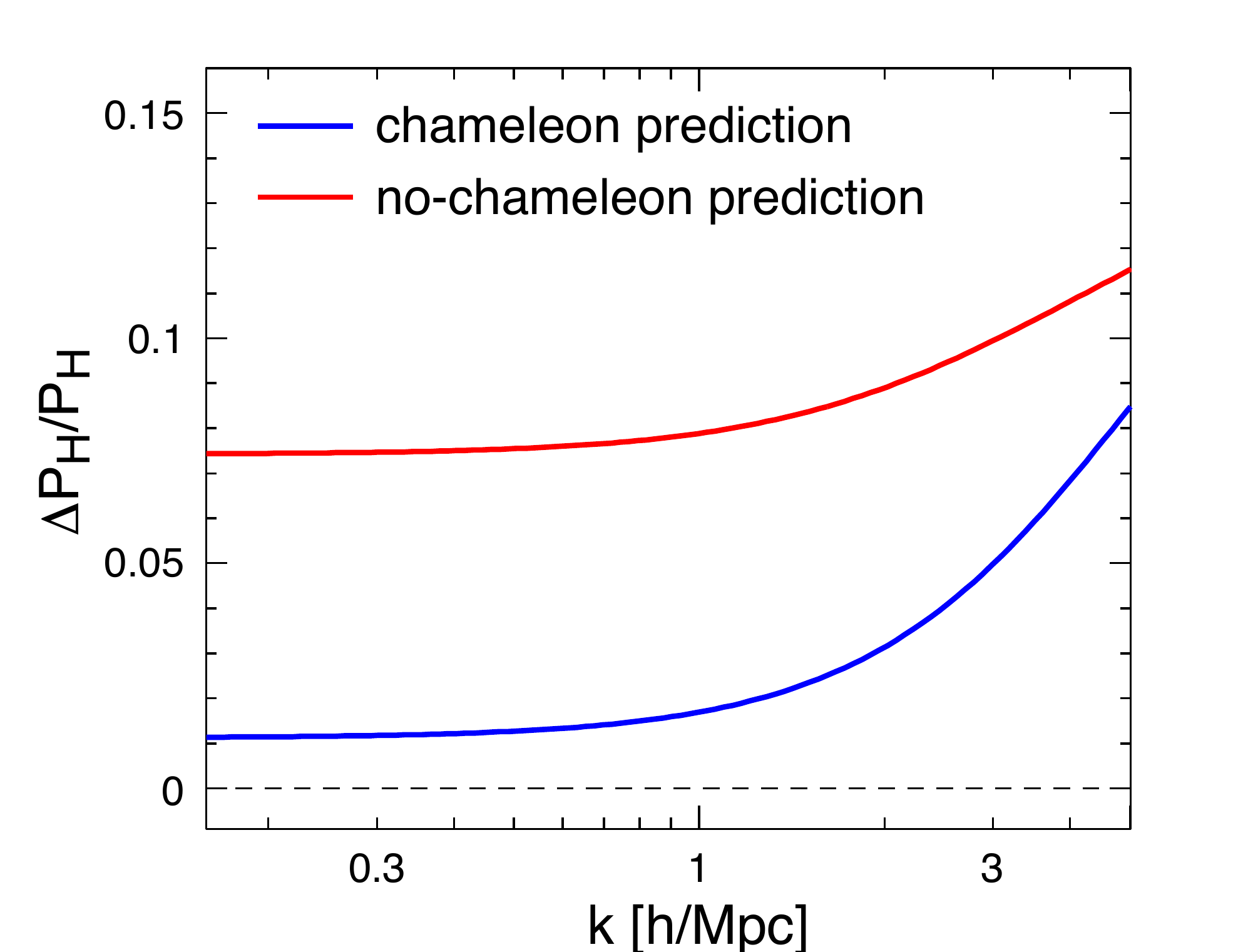}
\caption{One halo power spectrum enhancement of $f(R)$ over $\Lambda$CDM (dashed line).
 PPF predictions are shown for the $|f_{R0}|=10^{-6}$, $n=1$ model with (lower blue curve) 
and without (upper red curve) the chameleon modeling of the mass function effects.   
Without the chameleon modeling, high mass halos contribute excess power in the
quasilinear regime of  $k \lesssim$ few $h$/Mpc not found in the simulations.}
\label{fig:halo}
\end{figure}

As noted in \cite{Halopaper}, without a description of the chameleon mass function, the
one halo contributions for $|f_{R0}| < 10^{-5}$ 
are overestimated on intermediate scales where contributions
from groups and clusters dominate.  While this effect can be modeled with fitting parameters
that depend explicitly on the field value $|f_{R0}|$ \cite{Zhao:2010qy}, the halo model provides
a universal description of the chameleon effect on the power spectrum.   In addition it
provides better physical insight into its origin and relation to other observables
such as the mass function and higher point functions.  In Fig.~\ref{fig:halo}, we show the PPF predictions based
on the mass function enhancement for the one halo term compared with the no-chameleon
mass function predictions for $|f_{R0}|=10^{-6}$, $n=1$.  The chameleon effect suppresses
one halo power on scales less than a few $h$/Mpc.

For a full model of the power spectrum, we must include the large scale linear regime.
In the linear regime, the halo model describes the power spectrum in terms of the
correlation between two different halos.  Using a mass function construction that 
places all of the mass in halos and the halo bias
of the peak-background split guarantees that the two halo term simply returns the input linear
power spectrum.   
Unfortunately, in order to describe the power spectrum at intermediate $k$ between the linear and 
nonlinear regimes, the halo model requires complications such as halo exclusion to 
maintain accuracy (e.g.~\cite{Magliocchetti:2003ee,Tinker:2004gf,Cacciato:2008hm}).

\begin{figure}[htbp]
\centering
\includegraphics[width=0.5\textwidth]{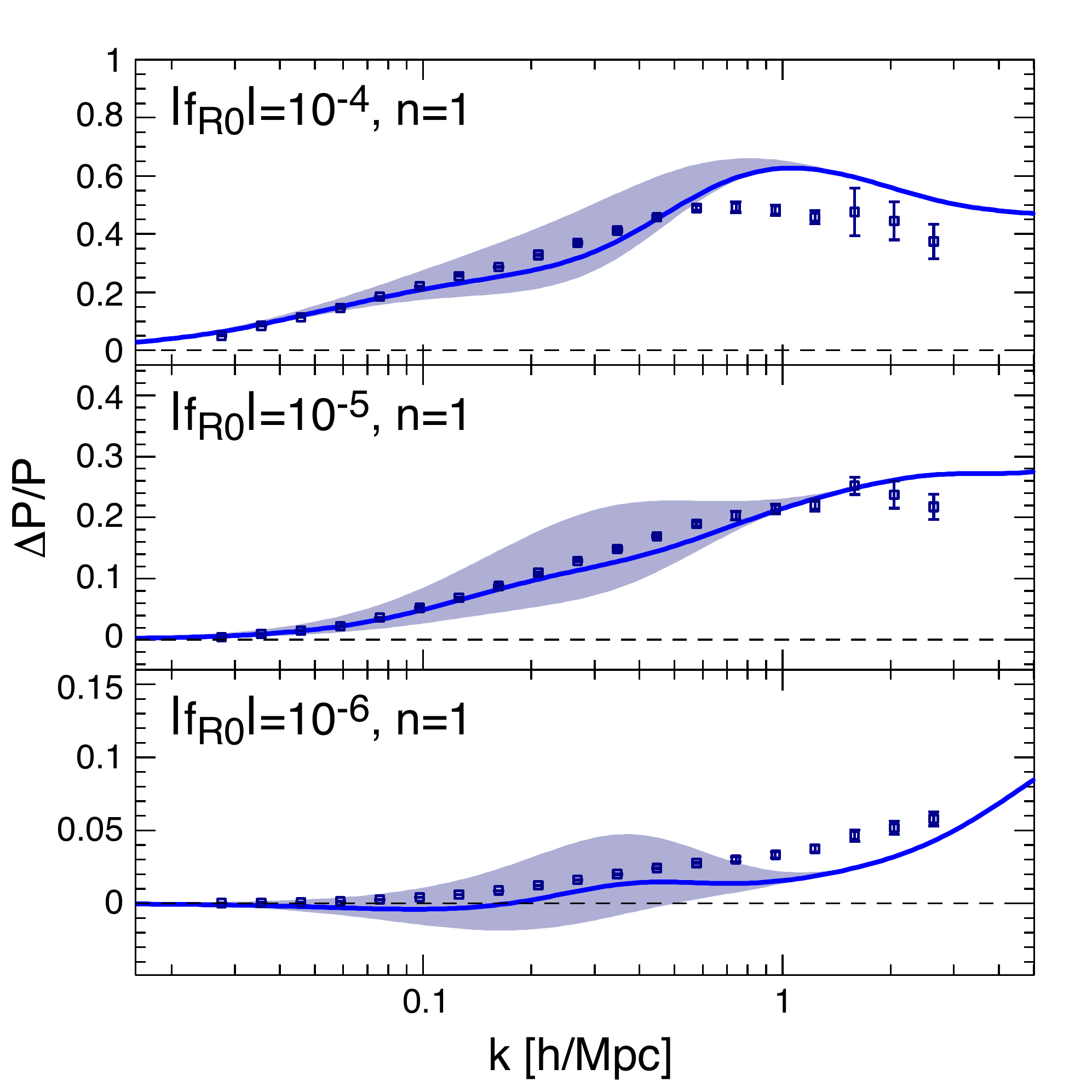}
\caption{Power spectrum enhancement for $n=1$ models compared with PPF predictions.  
Blue bands represent a range of Halofit parameters $\alpha_n$, $\beta_n$ where the
upper limit comes from the linear $\Lambda$CDM power spectrum and the
lower from the linear $f(R)$ power spectra.   Solid blue lines represent the average
of the two extreme values of the parameters which is a good prescription for all cases.}
\label{fig:Pk}
\end{figure}

We instead  take the phenomenological approach of Halofit \cite{smith03a} and seek
an interpolation between the known linear behavior and the one halo model for the deeply nonlinear
behavior
spectrum.  Specifically we take
\begin{equation}
{k^3 P(k) \over 2\pi^2}  = \Delta^2(k)=  \Delta^2_Q(k) + \Delta^2_H(k)\,.
\end{equation}
 Here $\Delta^2_H(k)$ is the dimensionless
power spectrum derived from $P_H(k)$ and $\Delta^2_Q(k)$ is related to the
linear theory power spectrum $\Delta^2_L(k)$ by
\begin{equation}
\Delta_Q^2(k) = \Delta^2_L { [1+\Delta_L^2(k)]^{\beta_n} \over 1 + \alpha_n \Delta_L^2(k)}
\exp(-y/4 - y^2/8)\,,
\end{equation}
where $y= k/k_\sigma$ determines the scale of the transition to the one halo term and
$\alpha_n$, $\beta_n$ are fitting parameters to adjust the shape of the transition.

Halofit describes the transition scale in terms of  the Gaussian filtered variance
\begin{equation}
\sigma_G^2(R) = \int d\ln k\,  \Delta_{\rm L}^2 e^{-k^2 R^2}
\end{equation}
as  $\sigma_G(k_\sigma^{-1})=1$. 
Rather than refit the transition parameters  $\alpha_n$, $\beta_n$, we examine the limiting
cases predicted by Halofit from the linear power spectra of $\Lambda$CDM and $f(R)$.
The Halofit prescription is to take the local slope %and curvature of the variance
\begin{eqnarray}
n_{\rm eff} &\equiv& -3 - {d \ln \sigma^2_G(R) \over d\ln R} \Big|_{\sigma_G=1},%\nonumber\\
%C &\equiv& -{d^2 \ln \sigma_G^2(R) \over d\ln R^2}  \Big|_{\sigma_G=1} .
\end{eqnarray}
and characterize
\begin{eqnarray}
\alpha_n &=& 1.3884 + 0.3700 n_{\rm eff} - 0.1452 n_{\rm eff}^2, \nonumber\\
\beta_n &=& 0.8291 + 0.9854 n_{\rm eff} + 0.3401 n_{\rm eff}^2.
\end{eqnarray}
In Fig.~\ref{fig:Pk} we show the simulation results compared with the PPF predictions
with a range of $\alpha_n$ and $\beta_n$ given by linear $\Lambda$CDM for the upper limit
and linear $f(R)$ for the lower.    Simulation results lie mainly in between these two limits
and in fact a simple average of the two $\alpha_n$ and $\beta_n$ values provides
a good description for all models.

%--------------
\section{Discussion}
 \label{sec:discussion}
%--------------

In the $f(R)$ model, the chameleon mechanism mediates a restoration of general relativity and ordinary Newtonian forces in deep gravitational potential wells.
We have shown that the main impact of the chameleon mechanism on cosmological statistics that depend on the
dark matter halo distribution can be simply described by a universal scaling for the
transition between modified and unmodified forces in mass or gravitational potential.   
This type of transition should be contrasted with phenomenological approaches
that implement the transition as a function of physical scale (e.g.~\cite{Amendola:2007rr}) or with models that implement
the Vainshtein mechanism where the transition is a function of density \cite{HuSaw07b}.

In the mass function, the chameleon mechanism leads to a doubly enhanced abundance
near the transition.  In the simulations, this enhancement is associated with smaller
mass halos still growing to the transition mass due to enhanced forces but transition
mass halos no longer merging into high mass halos due to the restoration of ordinary 
forces.   In our description, this mass conservation property is enforced through
 the Press-Schechter assumption
that all of the dark matter is in halos of some mass.   With this assumption, we can fit the
mass function results across a wide range in $f(R)$ models with two constants:
the scaling of the transition mass to the background field and the rapidity of the
transition in mass.

With a calibration of the mass function and the simulation result that halo profiles
as a function of mass are largely unchanged, we can construct predictions for the
$N$-point functions.   In particular for the power spectrum, including the chameleon
effects on the mass function brings predictions for the power spectrum excess down
in agreement with the simulations.   We have provided a simple modification to the
Halofit prescription to bridge the linear and nonlinear regimes.     For the higher $N$-point functions, 
the model predicts that the results  in the deeply nonlinear regime should scale mainly with
the single degree of freedom of the mass function rather than carry a unique signature
of $f(R)$ gravity \cite{GSHJV11}.  

Moreover, by describing these effects as enhancements over the $\Lambda$CDM mass function
and power spectrum with a physically well-motivated extrapolation to low masses and small scales
 rather than an absolute prediction of $f(R)$ statistics, one can use state-of-the-art simulations and
 mass calibrations for $\Lambda$CDM predictions and look for a parameterized excess over those
in the data (see, e.g.~\cite{Schmidt:2009am}).   Uncertainties in our $f(R)$ prescription merely translate into uncertainties in the $f(R)$ parameters rather than a false positive detection of modified gravity.    These techniques provide simple but approximate
means for testing the $f(R)$ model with weak lensing and other statistics that require large dynamic range and
precision.  Should in the future a positive detection occur, then high resolution gas simulations can
be performed to refine the calibration of cosmological observables in the $f(R)$ model.

 \smallskip
 \noindent {\it Acknowledgments}:   We thank Simone Ferraro, Surhud More and Fabian Schmidt for
 useful discussions. YL and WH were supported by
 the Kavli Institute for Cosmological Physics (KICP) at the University
 of Chicago through grants NSF PHY-0114422 and NSF PHY-0551142 and an
 endowment from the Kavli Foundation and its founder Fred Kavli.   WH
 was additionally supported by U.S.~Dept.\ of Energy contract
 DE-FG02-90ER-40560 and the David and Lucile Packard Foundation. 
  Computational resources for the cosmological simulations were provided by the KICP-Fermilab
computer cluster.

\vfill

%%%%%%%%%%%%%%%%%%%%%%%%%%%%%%%%%%%%%%%%%
\bibliography{ppffr}

%merlin.mbs apsrev4-1.bst 2010-07-25 4.21a (PWD, AO, DPC) hacked
%Control: key (0)
%Control: author (8) initials jnrlst
%Control: editor formatted (1) identically to author
%Control: production of article title (-1) disabled
%Control: page (0) single
%Control: year (1) truncated
%Control: production of eprint (0) enabled
\begin{thebibliography}{26}%
\makeatletter
\providecommand \@ifxundefined [1]{%
 \@ifx{#1\undefined}
}%
\providecommand \@ifnum [1]{%
 \ifnum #1\expandafter \@firstoftwo
 \else \expandafter \@secondoftwo
 \fi
}%
\providecommand \@ifx [1]{%
 \ifx #1\expandafter \@firstoftwo
 \else \expandafter \@secondoftwo
 \fi
}%
\providecommand \natexlab [1]{#1}%
\providecommand \enquote  [1]{``#1''}%
\providecommand \bibnamefont  [1]{#1}%
\providecommand \bibfnamefont [1]{#1}%
\providecommand \citenamefont [1]{#1}%
\providecommand \href@noop [0]{\@secondoftwo}%
\providecommand \href [0]{\begingroup \@sanitize@url \@href}%
\providecommand \@href[1]{\@@startlink{#1}\@@href}%
\providecommand \@@href[1]{\endgroup#1\@@endlink}%
\providecommand \@sanitize@url [0]{\catcode `\\12\catcode `\$12\catcode
  `\&12\catcode `\#12\catcode `\^12\catcode `\_12\catcode `\%12\relax}%
\providecommand \@@startlink[1]{}%
\providecommand \@@endlink[0]{}%
\providecommand \url  [0]{\begingroup\@sanitize@url \@url }%
\providecommand \@url [1]{\endgroup\@href {#1}{\urlprefix }}%
\providecommand \urlprefix  [0]{URL }%
\providecommand \Eprint [0]{\href }%
\providecommand \doibase [0]{http://dx.doi.org/}%
\providecommand \selectlanguage [0]{\@gobble}%
\providecommand \bibinfo  [0]{\@secondoftwo}%
\providecommand \bibfield  [0]{\@secondoftwo}%
\providecommand \translation [1]{[#1]}%
\providecommand \BibitemOpen [0]{}%
\providecommand \bibitemStop [0]{}%
\providecommand \bibitemNoStop [0]{.\EOS\space}%
\providecommand \EOS [0]{\spacefactor3000\relax}%
\providecommand \BibitemShut  [1]{\csname bibitem#1\endcsname}%
\let\auto@bib@innerbib\@empty
%</preamble>
\bibitem [{\citenamefont {Capozziello}(2002)}]{Capozziello:2002rd}%
  \BibitemOpen
  \bibfield  {author} {\bibinfo {author} {\bibfnamefont {S.}~\bibnamefont
  {Capozziello}},\ }\href {\doibase 10.1142/S0218271802002025} {\bibfield
  {journal} {\bibinfo  {journal} {Int.J.Mod.Phys.}\ }\textbf {\bibinfo {volume}
  {D11}},\ \bibinfo {pages} {483} (\bibinfo {year} {2002})},\ \Eprint
  {http://arxiv.org/abs/gr-qc/0201033} {arXiv:gr-qc/0201033 [gr-qc]}
  \BibitemShut {NoStop}%
\bibitem [{\citenamefont {Nojiri}\ and\ \citenamefont
  {Odintsov}(2003)}]{NojOdi03}%
  \BibitemOpen
  \bibfield  {author} {\bibinfo {author} {\bibfnamefont {S.}~\bibnamefont
  {Nojiri}}\ and\ \bibinfo {author} {\bibfnamefont {S.~D.}\ \bibnamefont
  {Odintsov}},\ }\href@noop {} {\bibfield  {journal} {\bibinfo  {journal}
  {Phys. Rev.}\ }\textbf {\bibinfo {volume} {D68}},\ \bibinfo {pages} {123512}
  (\bibinfo {year} {2003})},\ \Eprint {http://arxiv.org/abs/hep-th/0307288}
  {hep-th/0307288} \BibitemShut {NoStop}%
%%CITATION = HEP-TH 0307288;%%
\bibitem [{\citenamefont {Carroll}\ \emph {et~al.}(2004)\citenamefont
  {Carroll}, \citenamefont {Duvvuri}, \citenamefont {Trodden},\ and\
  \citenamefont {Turner}}]{Caretal03}%
  \BibitemOpen
  \bibfield  {author} {\bibinfo {author} {\bibfnamefont {S.~M.}\ \bibnamefont
  {Carroll}}, \bibinfo {author} {\bibfnamefont {V.}~\bibnamefont {Duvvuri}},
  \bibinfo {author} {\bibfnamefont {M.}~\bibnamefont {Trodden}}, \ and\
  \bibinfo {author} {\bibfnamefont {M.~S.}\ \bibnamefont {Turner}},\
  }\href@noop {} {\bibfield  {journal} {\bibinfo  {journal} {Phys. Rev.}\
  }\textbf {\bibinfo {volume} {D70}},\ \bibinfo {pages} {043528} (\bibinfo
  {year} {2004})},\ \Eprint {http://arxiv.org/abs/astro-ph/0306438}
  {astro-ph/0306438} \BibitemShut {NoStop}%
%%CITATION = ASTRO-PH 0306438;%%
\bibitem [{\citenamefont {Mota}\ and\ \citenamefont
  {Barrow}(2004)}]{Mota:2003tc}%
  \BibitemOpen
  \bibfield  {author} {\bibinfo {author} {\bibfnamefont {D.~F.}\ \bibnamefont
  {Mota}}\ and\ \bibinfo {author} {\bibfnamefont {J.~D.}\ \bibnamefont
  {Barrow}},\ }\href {\doibase 10.1016/j.physletb.2003.12.016} {\bibfield
  {journal} {\bibinfo  {journal} {Phys. Lett.}\ }\textbf {\bibinfo {volume}
  {B581}},\ \bibinfo {pages} {141} (\bibinfo {year} {2004})},\ \Eprint
  {http://arxiv.org/abs/astro-ph/0306047} {arXiv:astro-ph/0306047} \BibitemShut
  {NoStop}%
%%CITATION = ASTRO-PH/0306047;%%
\bibitem [{\citenamefont {{Khoury}}\ and\ \citenamefont
  {{Weltman}}(2004)}]{khoury04a}%
  \BibitemOpen
  \bibfield  {author} {\bibinfo {author} {\bibfnamefont {J.}~\bibnamefont
  {{Khoury}}}\ and\ \bibinfo {author} {\bibfnamefont {A.}~\bibnamefont
  {{Weltman}}},\ }\href@noop {} {\bibfield  {journal} {\bibinfo  {journal}
  {\prd}\ }\textbf {\bibinfo {volume} {69}},\ \bibinfo {pages} {044026}
  (\bibinfo {year} {2004})},\ \Eprint
  {http://arxiv.org/abs/arXiv:astro-ph/0309411} {arXiv:astro-ph/0309411}
  \BibitemShut {NoStop}%
\bibitem [{\citenamefont {{Hu}}\ and\ \citenamefont
  {{Sawicki}}(2007{\natexlab{a}})}]{HuSaw07a}%
  \BibitemOpen
  \bibfield  {author} {\bibinfo {author} {\bibfnamefont {W.}~\bibnamefont
  {{Hu}}}\ and\ \bibinfo {author} {\bibfnamefont {I.}~\bibnamefont
  {{Sawicki}}},\ }\href {\doibase 10.1103/PhysRevD.76.064004} {\bibfield
  {journal} {\bibinfo  {journal} {\prd}\ }\textbf {\bibinfo {volume} {76}},\
  \bibinfo {pages} {064004} (\bibinfo {year} {2007}{\natexlab{a}})},\ \Eprint
  {http://arxiv.org/abs/arXiv:0705.1158} {arXiv:0705.1158} \BibitemShut
  {NoStop}%
\bibitem [{\citenamefont {Schmidt}\ \emph
  {et~al.}(2009{\natexlab{a}})\citenamefont {Schmidt}, \citenamefont
  {Vikhlinin},\ and\ \citenamefont {Hu}}]{Schmidt:2009am}%
  \BibitemOpen
  \bibfield  {author} {\bibinfo {author} {\bibfnamefont {F.}~\bibnamefont
  {Schmidt}}, \bibinfo {author} {\bibfnamefont {A.}~\bibnamefont {Vikhlinin}},
  \ and\ \bibinfo {author} {\bibfnamefont {W.}~\bibnamefont {Hu}},\ }\href
  {\doibase 10.1103/PhysRevD.80.083505} {\bibfield  {journal} {\bibinfo
  {journal} {Phys. Rev.}\ }\textbf {\bibinfo {volume} {D80}},\ \bibinfo {pages}
  {083505} (\bibinfo {year} {2009}{\natexlab{a}})},\ \Eprint
  {http://arxiv.org/abs/0908.2457} {arXiv:0908.2457 [astro-ph.CO]} \BibitemShut
  {NoStop}%
%%CITATION = 0908.2457;%%
\bibitem [{\citenamefont {Lombriser}\ \emph {et~al.}(2010)\citenamefont
  {Lombriser}, \citenamefont {Slosar}, \citenamefont {Seljak},\ and\
  \citenamefont {Hu}}]{Lombriser:2010mp}%
  \BibitemOpen
  \bibfield  {author} {\bibinfo {author} {\bibfnamefont {L.}~\bibnamefont
  {Lombriser}}, \bibinfo {author} {\bibfnamefont {A.}~\bibnamefont {Slosar}},
  \bibinfo {author} {\bibfnamefont {U.}~\bibnamefont {Seljak}}, \ and\ \bibinfo
  {author} {\bibfnamefont {W.}~\bibnamefont {Hu}},\ }\href@noop {} {\
  (\bibinfo {year} {2010})},\ \Eprint {http://arxiv.org/abs/1003.3009}
  {arXiv:1003.3009 [astro-ph.CO]} \BibitemShut {NoStop}%
%%CITATION = 1003.3009;%%
\bibitem [{\citenamefont {Ferraro}\ \emph {et~al.}(2011)\citenamefont
  {Ferraro}, \citenamefont {Schmidt},\ and\ \citenamefont
  {Hu}}]{Ferraro:2010gh}%
  \BibitemOpen
  \bibfield  {author} {\bibinfo {author} {\bibfnamefont {S.}~\bibnamefont
  {Ferraro}}, \bibinfo {author} {\bibfnamefont {F.}~\bibnamefont {Schmidt}}, \
  and\ \bibinfo {author} {\bibfnamefont {W.}~\bibnamefont {Hu}},\ }\href
  {\doibase 10.1103/PhysRevD.83.063503} {\bibfield  {journal} {\bibinfo
  {journal} {Phys.Rev.}\ }\textbf {\bibinfo {volume} {D83}},\ \bibinfo {pages}
  {063503} (\bibinfo {year} {2011})},\ \Eprint {http://arxiv.org/abs/1011.0992}
  {arXiv:1011.0992 [astro-ph.CO]} \BibitemShut {NoStop}%
\bibitem [{\citenamefont {Schmidt}\ \emph
  {et~al.}(2009{\natexlab{b}})\citenamefont {Schmidt}, \citenamefont {Lima},
  \citenamefont {Oyaizu},\ and\ \citenamefont {Hu}}]{Halopaper}%
  \BibitemOpen
  \bibfield  {author} {\bibinfo {author} {\bibfnamefont {F.}~\bibnamefont
  {Schmidt}}, \bibinfo {author} {\bibfnamefont {M.~V.}\ \bibnamefont {Lima}},
  \bibinfo {author} {\bibfnamefont {H.}~\bibnamefont {Oyaizu}}, \ and\ \bibinfo
  {author} {\bibfnamefont {W.}~\bibnamefont {Hu}},\ }\href {\doibase
  10.1103/PhysRevD.79.083518} {\bibfield  {journal} {\bibinfo  {journal} {Phys.
  Rev.}\ }\textbf {\bibinfo {volume} {D79}},\ \bibinfo {pages} {083518}
  (\bibinfo {year} {2009}{\natexlab{b}})},\ \Eprint
  {http://arxiv.org/abs/0812.0545} {arXiv:0812.0545 [astro-ph]} \BibitemShut
  {NoStop}%
%%CITATION = 0812.0545;%%
\bibitem [{\citenamefont {{Hu}}\ and\ \citenamefont
  {{Sawicki}}(2007{\natexlab{b}})}]{HuSaw07b}%
  \BibitemOpen
  \bibfield  {author} {\bibinfo {author} {\bibfnamefont {W.}~\bibnamefont
  {{Hu}}}\ and\ \bibinfo {author} {\bibfnamefont {I.}~\bibnamefont
  {{Sawicki}}},\ }\href {\doibase 10.1103/PhysRevD.76.104043} {\bibfield
  {journal} {\bibinfo  {journal} {\prd}\ }\textbf {\bibinfo {volume} {76}},\
  \bibinfo {pages} {104043} (\bibinfo {year} {2007}{\natexlab{b}})},\ \Eprint
  {http://arxiv.org/abs/arXiv:0708.1190} {arXiv:0708.1190} \BibitemShut
  {NoStop}%
\bibitem [{\citenamefont {Koyama}\ \emph {et~al.}(2009)\citenamefont {Koyama},
  \citenamefont {Taruya},\ and\ \citenamefont {Hiramatsu}}]{Koyama:2009me}%
  \BibitemOpen
  \bibfield  {author} {\bibinfo {author} {\bibfnamefont {K.}~\bibnamefont
  {Koyama}}, \bibinfo {author} {\bibfnamefont {A.}~\bibnamefont {Taruya}}, \
  and\ \bibinfo {author} {\bibfnamefont {T.}~\bibnamefont {Hiramatsu}},\ }\href
  {\doibase 10.1103/PhysRevD.79.123512} {\bibfield  {journal} {\bibinfo
  {journal} {Phys. Rev.}\ }\textbf {\bibinfo {volume} {D79}},\ \bibinfo {pages}
  {123512} (\bibinfo {year} {2009})},\ \Eprint {http://arxiv.org/abs/0902.0618}
  {arXiv:0902.0618 [astro-ph.CO]} \BibitemShut {NoStop}%
%%CITATION = 0902.0618;%%
\bibitem [{\citenamefont {Oyaizu}\ \emph {et~al.}(2008)\citenamefont {Oyaizu},
  \citenamefont {Lima},\ and\ \citenamefont {Hu}}]{Pkpaper}%
  \BibitemOpen
  \bibfield  {author} {\bibinfo {author} {\bibfnamefont {H.}~\bibnamefont
  {Oyaizu}}, \bibinfo {author} {\bibfnamefont {M.}~\bibnamefont {Lima}}, \ and\
  \bibinfo {author} {\bibfnamefont {W.}~\bibnamefont {Hu}},\ }\href {\doibase
  10.1103/PhysRevD.78.123524} {\bibfield  {journal} {\bibinfo  {journal} {Phys.
  Rev.}\ }\textbf {\bibinfo {volume} {D78}},\ \bibinfo {pages} {123524}
  (\bibinfo {year} {2008})},\ \Eprint {http://arxiv.org/abs/0807.2462}
  {arXiv:0807.2462 [astro-ph]} \BibitemShut {NoStop}%
%%CITATION = 0807.2462;%%
\bibitem [{Note1()}]{Note1}%
  \BibitemOpen
  \bibinfo {note} {This corrects a 3\% absolute error in the initial amplitude
  reported in \cite {Pkpaper} but does not change results relative to $\Lambda
  $CDM significantly.}\BibitemShut {Stop}%
\bibitem [{\citenamefont {Tinker}\ \emph {et~al.}(2008)\citenamefont {Tinker}
  \emph {et~al.}}]{Tin08}%
  \BibitemOpen
  \bibfield  {author} {\bibinfo {author} {\bibfnamefont {J.~L.}\ \bibnamefont
  {Tinker}} \emph {et~al.},\ }\href {\doibase 10.1086/591439} {\bibfield
  {journal} {\bibinfo  {journal} {Astrophys. J.}\ }\textbf {\bibinfo {volume}
  {688}},\ \bibinfo {pages} {709} (\bibinfo {year} {2008})},\ \Eprint
  {http://arxiv.org/abs/0803.2706} {arXiv:0803.2706 [astro-ph]} \BibitemShut
  {NoStop}%
%%CITATION = 0803.2706;%%
\bibitem [{\citenamefont {{Lacey}}\ and\ \citenamefont
  {{Cole}}(1994)}]{LacCol94}%
  \BibitemOpen
  \bibfield  {author} {\bibinfo {author} {\bibfnamefont {C.}~\bibnamefont
  {{Lacey}}}\ and\ \bibinfo {author} {\bibfnamefont {S.}~\bibnamefont
  {{Cole}}},\ }\href@noop {} {\bibfield  {journal} {\bibinfo  {journal}
  {\mnras}\ }\textbf {\bibinfo {volume} {271}},\ \bibinfo {pages} {676}
  (\bibinfo {year} {1994})},\ \Eprint
  {http://arxiv.org/abs/arXiv:astro-ph/9402069} {arXiv:astro-ph/9402069}
  \BibitemShut {NoStop}%
\bibitem [{\citenamefont {{Sheth}}\ and\ \citenamefont
  {{Tormen}}(1999)}]{SheTor99}%
  \BibitemOpen
  \bibfield  {author} {\bibinfo {author} {\bibfnamefont {R.}~\bibnamefont
  {{Sheth}}}\ and\ \bibinfo {author} {\bibfnamefont {B.}~\bibnamefont
  {{Tormen}}},\ }\href@noop {} {\bibfield  {journal} {\bibinfo  {journal}
  {\mnras}\ }\textbf {\bibinfo {volume} {308}},\ \bibinfo {pages} {119}
  (\bibinfo {year} {1999})}\BibitemShut {NoStop}%
\bibitem [{\citenamefont {{Hu}}\ and\ \citenamefont
  {{Kravtsov}}(2003)}]{HuKravtsov}%
  \BibitemOpen
  \bibfield  {author} {\bibinfo {author} {\bibfnamefont {W.}~\bibnamefont
  {{Hu}}}\ and\ \bibinfo {author} {\bibfnamefont {A.~V.}\ \bibnamefont
  {{Kravtsov}}},\ }\href {\doibase 10.1086/345846} {\bibfield  {journal}
  {\bibinfo  {journal} {\apj}\ }\textbf {\bibinfo {volume} {584}},\ \bibinfo
  {pages} {702} (\bibinfo {year} {2003})},\ \Eprint
  {http://arxiv.org/abs/arXiv:astro-ph/0203169} {arXiv:astro-ph/0203169}
  \BibitemShut {NoStop}%
\bibitem [{\citenamefont {{Bullock}}\ \emph {et~al.}(2001)\citenamefont
  {{Bullock}}, \citenamefont {{Kolatt}}, \citenamefont {{Sigad}}, \citenamefont
  {{Somerville}}, \citenamefont {{Kravtsov}}, \citenamefont {{Klypin}},
  \citenamefont {{Primack}},\ and\ \citenamefont {{Dekel}}}]{Buletal01}%
  \BibitemOpen
  \bibfield  {author} {\bibinfo {author} {\bibfnamefont {J.~S.}\ \bibnamefont
  {{Bullock}}}, \bibinfo {author} {\bibfnamefont {T.~S.}\ \bibnamefont
  {{Kolatt}}}, \bibinfo {author} {\bibfnamefont {Y.}~\bibnamefont {{Sigad}}},
  \bibinfo {author} {\bibfnamefont {R.~S.}\ \bibnamefont {{Somerville}}},
  \bibinfo {author} {\bibfnamefont {A.~V.}\ \bibnamefont {{Kravtsov}}},
  \bibinfo {author} {\bibfnamefont {A.~A.}\ \bibnamefont {{Klypin}}}, \bibinfo
  {author} {\bibfnamefont {J.~R.}\ \bibnamefont {{Primack}}}, \ and\ \bibinfo
  {author} {\bibfnamefont {A.}~\bibnamefont {{Dekel}}},\ }\href@noop {}
  {\bibfield  {journal} {\bibinfo  {journal} {\mnras}\ }\textbf {\bibinfo
  {volume} {321}},\ \bibinfo {pages} {559} (\bibinfo {year} {2001})},\ \Eprint
  {http://arxiv.org/abs/arXiv:astro-ph/9908159} {arXiv:astro-ph/9908159}
  \BibitemShut {NoStop}%
\bibitem [{\citenamefont {Zhao}\ \emph {et~al.}(2011)\citenamefont {Zhao},
  \citenamefont {Li},\ and\ \citenamefont {Koyama}}]{Zhao:2010qy}%
  \BibitemOpen
  \bibfield  {author} {\bibinfo {author} {\bibfnamefont {G.-B.}\ \bibnamefont
  {Zhao}}, \bibinfo {author} {\bibfnamefont {B.}~\bibnamefont {Li}}, \ and\
  \bibinfo {author} {\bibfnamefont {K.}~\bibnamefont {Koyama}},\ }\href
  {\doibase 10.1103/PhysRevD.83.044007} {\bibfield  {journal} {\bibinfo
  {journal} {Phys.Rev.}\ }\textbf {\bibinfo {volume} {D83}},\ \bibinfo {pages}
  {044007} (\bibinfo {year} {2011})},\ \Eprint {http://arxiv.org/abs/1011.1257}
  {arXiv:1011.1257 [astro-ph.CO]} \BibitemShut {NoStop}%
\bibitem [{\citenamefont {Magliocchetti}\ and\ \citenamefont
  {Porciani}(2003)}]{Magliocchetti:2003ee}%
  \BibitemOpen
  \bibfield  {author} {\bibinfo {author} {\bibfnamefont {M.}~\bibnamefont
  {Magliocchetti}}\ and\ \bibinfo {author} {\bibfnamefont {C.}~\bibnamefont
  {Porciani}},\ }\href {\doibase 10.1046/j.1365-2966.2003.07094.x} {\bibfield
  {journal} {\bibinfo  {journal} {Mon.Not.Roy.Astron.Soc.}\ }\textbf {\bibinfo
  {volume} {346}},\ \bibinfo {pages} {186} (\bibinfo {year} {2003})},\ \Eprint
  {http://arxiv.org/abs/astro-ph/0304003} {arXiv:astro-ph/0304003 [astro-ph]}
  \BibitemShut {NoStop}%
\bibitem [{\citenamefont {Tinker}\ \emph {et~al.}(2005)\citenamefont {Tinker},
  \citenamefont {Weinberg}, \citenamefont {Zheng},\ and\ \citenamefont
  {Zehavi}}]{Tinker:2004gf}%
  \BibitemOpen
  \bibfield  {author} {\bibinfo {author} {\bibfnamefont {J.~L.}\ \bibnamefont
  {Tinker}}, \bibinfo {author} {\bibfnamefont {D.~H.}\ \bibnamefont
  {Weinberg}}, \bibinfo {author} {\bibfnamefont {Z.}~\bibnamefont {Zheng}}, \
  and\ \bibinfo {author} {\bibfnamefont {I.}~\bibnamefont {Zehavi}},\ }\href
  {\doibase 10.1086/432084} {\bibfield  {journal} {\bibinfo  {journal}
  {Astrophys.J.}\ }\textbf {\bibinfo {volume} {631}},\ \bibinfo {pages} {41}
  (\bibinfo {year} {2005})},\ \Eprint {http://arxiv.org/abs/astro-ph/0411777}
  {arXiv:astro-ph/0411777 [astro-ph]} \BibitemShut {NoStop}%
\bibitem [{\citenamefont {Cacciato}\ \emph {et~al.}(2009)\citenamefont
  {Cacciato}, \citenamefont {Bosch}, \citenamefont {More}, \citenamefont {Li},
  \citenamefont {Mo} \emph {et~al.}}]{Cacciato:2008hm}%
  \BibitemOpen
  \bibfield  {author} {\bibinfo {author} {\bibfnamefont {M.}~\bibnamefont
  {Cacciato}}, \bibinfo {author} {\bibfnamefont {F.~C.~d.}\ \bibnamefont
  {Bosch}}, \bibinfo {author} {\bibfnamefont {S.}~\bibnamefont {More}},
  \bibinfo {author} {\bibfnamefont {R.}~\bibnamefont {Li}}, \bibinfo {author}
  {\bibfnamefont {H.}~\bibnamefont {Mo}},  \emph {et~al.},\ }\href {\doibase
  10.1111/j.1365-2966.2008.14362.x} {\bibfield  {journal} {\bibinfo  {journal}
  {Mon.Not.Roy.Astron.Soc.}\ }\textbf {\bibinfo {volume} {394}},\ \bibinfo
  {pages} {929} (\bibinfo {year} {2009})},\ \Eprint
  {http://arxiv.org/abs/0807.4932} {arXiv:0807.4932 [astro-ph]} \BibitemShut
  {NoStop}%
\bibitem [{\citenamefont {{Smith}}\ \emph {et~al.}(2003)\citenamefont
  {{Smith}}, \citenamefont {{Peacock}}, \citenamefont {{Jenkins}},
  \citenamefont {{White}}, \citenamefont {{Frenk}}, \citenamefont {{Pearce}},
  \citenamefont {{Thomas}}, \citenamefont {{Efstathiou}},\ and\ \citenamefont
  {{Couchman}}}]{smith03a}%
  \BibitemOpen
  \bibfield  {author} {\bibinfo {author} {\bibfnamefont {R.~E.}\ \bibnamefont
  {{Smith}}}, \bibinfo {author} {\bibfnamefont {J.~A.}\ \bibnamefont
  {{Peacock}}}, \bibinfo {author} {\bibfnamefont {A.}~\bibnamefont
  {{Jenkins}}}, \bibinfo {author} {\bibfnamefont {S.~D.~M.}\ \bibnamefont
  {{White}}}, \bibinfo {author} {\bibfnamefont {C.~S.}\ \bibnamefont
  {{Frenk}}}, \bibinfo {author} {\bibfnamefont {F.~R.}\ \bibnamefont
  {{Pearce}}}, \bibinfo {author} {\bibfnamefont {P.~A.}\ \bibnamefont
  {{Thomas}}}, \bibinfo {author} {\bibfnamefont {G.}~\bibnamefont
  {{Efstathiou}}}, \ and\ \bibinfo {author} {\bibfnamefont {H.~M.~P.}\
  \bibnamefont {{Couchman}}},\ }\href {\doibase
  10.1046/j.1365-8711.2003.06503.x} {\bibfield  {journal} {\bibinfo  {journal}
  {\mnras}\ }\textbf {\bibinfo {volume} {341}},\ \bibinfo {pages} {1311}
  (\bibinfo {year} {2003})},\ \Eprint
  {http://arxiv.org/abs/arXiv:astro-ph/0207664} {arXiv:astro-ph/0207664}
  \BibitemShut {NoStop}%
\bibitem [{\citenamefont {Amendola}\ \emph {et~al.}(2008)\citenamefont
  {Amendola}, \citenamefont {Kunz},\ and\ \citenamefont
  {Sapone}}]{Amendola:2007rr}%
  \BibitemOpen
  \bibfield  {author} {\bibinfo {author} {\bibfnamefont {L.}~\bibnamefont
  {Amendola}}, \bibinfo {author} {\bibfnamefont {M.}~\bibnamefont {Kunz}}, \
  and\ \bibinfo {author} {\bibfnamefont {D.}~\bibnamefont {Sapone}},\ }\href
  {\doibase 10.1088/1475-7516/2008/04/013} {\bibfield  {journal} {\bibinfo
  {journal} {JCAP}\ }\textbf {\bibinfo {volume} {0804}},\ \bibinfo {pages}
  {013} (\bibinfo {year} {2008})},\ \Eprint {http://arxiv.org/abs/0704.2421}
  {arXiv:0704.2421 [astro-ph]} \BibitemShut {NoStop}%
\bibitem [{\citenamefont {Gil-Marin}\ \emph {et~al.}(2011)\citenamefont
  {Gil-Marin}, \citenamefont {Schmidt}, \citenamefont {Hu}, \citenamefont
  {Jimenez},\ and\ \citenamefont {Verde}}]{GSHJV11}%
  \BibitemOpen
  \bibfield  {author} {\bibinfo {author} {\bibfnamefont {H.}~\bibnamefont
  {Gil-Marin}}, \bibinfo {author} {\bibfnamefont {F.}~\bibnamefont {Schmidt}},
  \bibinfo {author} {\bibfnamefont {W.}~\bibnamefont {Hu}}, \bibinfo {author}
  {\bibfnamefont {R.}~\bibnamefont {Jimenez}}, \ and\ \bibinfo {author}
  {\bibfnamefont {L.}~\bibnamefont {Verde}},\ }\href@noop {} {\bibfield
  {journal} {\bibinfo  {journal} {JCAP}\ }\textbf {\bibinfo {volume} {\rm
  submitted}} (\bibinfo {year} {2011})},\ \Eprint
  {http://arxiv.org/abs/1109.2115} {arXiv:1109.2115} \BibitemShut {NoStop}%
\end{thebibliography}%
%%%%%%%%%%%%%%%%%%%%%%%%%%%%%%%%%%%%%%%%%

\end{document}